\documentstyle[12pt,epsfig,amsmath,amssymb]{article}
\begin{document}

\title{\bf Computational costs of data definition at the quantum - classical interface}

\author{{Chris Fields}\\ \\
{\it 21 Rue des Lavandi\`eres}\\
{\it Caunes Minervois 11160 France}\\ \\
{fieldsres@gmail.com}}
\maketitle

\begin{abstract}
Model-independent semantic requirements for user specification and interpretation of data before and after quantum computations are characterized.  Classical computational costs of assigning classical data values to quantum registers and to run-time parameters passed across a classical-to-quantum application programming interface are derived.  It is shown that the classical computational costs of data definition equal or exceed the classical computational cost of solving the problem of interest for all applications of quantum computing except computations defined over the integers and the simulation of linear systems with linear boundary conditions.
\end{abstract}


\section{Introduction}

All computations start and end at the user-interface (UI) level of a macroscopic physical device.  Prior to initiating a computation, a user provides data that represent actual or possible states of the world as input.  At the end of the computation, the user collects data representing actual or possible states of the world as output.  Thus a fundamental requirement of the encoding of data by any physical device that executes computations is that the user be able to directly interpret the UI level encodings of both input and output.  This fundamental semantic requirement at the UI level drives the interpretation of the physical behavior of devices that function as computers in terms of virtual machines that perform meaningful abstract operations on meaningful data \cite{tan76, smith05}.  As Aaronson has pointed out in a somewhat different context \cite{aaronson05}, without the possibility of such an interpretation, the intuitive notion of ``computation'' collapses.

This paper explores the consequences of the requirement of UI level, or equivalently application-programming interface (API) level interpretability for the design and implementation of data interfaces between classical and quantum processing elements of a hybrid computational device.  It has two goals.  The first is to characterize, independently of the model of quantum computation, the functional requirements for the data interfaces that define the classical-side referents of the global states of qubit registers and of the run-time parameters employed to specify the behavior of quantum algorithms that act on those qubit registers.  The second goal is to derive from these functional requirements the classical computational costs of data definition at the quantum-classical interface.  It is shown that the classical computational costs of UI or API level data definition equal or exceed the classical computational cost of solving the problem of interest for all applications of quantum computing except computations defined over the integers and the simulation of linear systems with linear boundary conditions.

\section{Notation}

Let $\mathbf{D}$ be a physical device, the time evolution of which is described by a Hamiltonian $\mathcal{H}$.  Let $|\mathbf{D}>|_{t}$ denote the state of $\mathbf{D}$ at a particular time $t$, so that the action of $\mathcal{H}$ over a time interval $\Delta t$ can be written:

\begin{equation}
\mathcal{H}: |\mathbf{D}>|_{t} \longmapsto |\mathbf{D}>|_{t + \Delta t}.
\end{equation}

A \textit{component} $\mathbf{C}$ of $\mathbf{D}$ is a subsystem comprising a proper subset of degrees of freedom of $\mathbf{D}$; the state of a component at time $t_{i}$ can be written in terms of the projection $|\mathbf{C}><\mathbf{C}|$ as $|\mathbf{C}>|_{t_{i}} = |\mathbf{C}><\mathbf{C}||\mathbf{D}>|_{t_{i}}$.  A quantum register $\mathbf{R}$ is a component of $\mathbf{D}$ in this dynamic sense.  The local Hamiltonian $|\mathbf{R}><\mathbf{R}|\mathcal{H}$ acting on a suitably-isolated quantum register can be interpreted as implementing a quantum algorithm $\mathbf{Q}$ defined over $\mathbf{R}$.  It will be assumed that $\mathbf{R}$ comprises a linear array of $n$ physical devices, each of which implements a qubit; the physical address of the $i^{th}$ qubit will be taken to be the integer $i$.  Qubit states will be written in the computational basis $\{|0>, |1>\}$; global states of $\mathbf{R}$ in the computational basis will for convenience be denoted $|1>, |2>, ..., |N>$.  

Users interact directly not with quantum states of $\mathbf{D}$, but with re-identifiable macroscopic states of $\mathbf{D}$.  In order to avoid imposing a particular interpretation of quantum measurement and hence a particular definition of ``macroscopic state", such macroscopic states are here considered not as measured physical states of $\mathbf{D}$, but rather as states of a virtual machine emulated by $\mathbf{D}$.  In particular, let $\{D^{k}_{i}\}$ denote a set of virtual states at a level of abstraction $k$, and let $E^{k}$ be an emulation mapping:

\begin{equation}
E^{k}: \{|\mathbf{D}_{\mathit{i}}>\}^{\mathit{k}}_{\mathit{j}} \longmapsto \mathit{D^{k}_{j}}
\end{equation}

that maps an equivalence class $\{|\mathbf{D}_{\mathit{i}}>\}^{\mathit{k}}_{\mathit{j}}$ of $E^{k}$-\textit{indistinguishable} states of $\mathbf{D}$ onto the $j^{th}$ virtual state $D^{k}_{j}$ at the $k^{th}$ level of abstraction.  Two such levels of abstraction will be of concern here: the \textit{classical hardware level} comprising a finite set of virtual states $\{D^{H}_{j}\}$ produced by an emulation mapping $E^{H}$ and the \textit{user-interface level} comprising a finite set of virtual states $\{D^{UI}_{j}\}$ produced by an emulation mapping $E^{UI}$.  The virtual states at both of these levels of abstraction are taken to have dimensionality much smaller than that of $\mathbf{D}$; hence each equivalence class $\{|\mathbf{D}_{\mathit{i}}>\}^{\mathit{k}}$ is non-trivial.

Treating the classical hardware level of $\mathbf{D}$ as a virtual machine is unusual.  Doing so is consistent, however, with the usual motivations for defining a virtual machine.  The classical hardware level is canonically characterized in terms of bits implemented by bistable devices, the underlying physics of which is typically ignored.  Treating these bits as ``physical" is in effect a shorthand: it indicates that no lower-level interpretation of the device is relevant to its computational characterization.  In the case of a quantum computer, however, a lower-level interpretation of at least some components of the device clearly is relevant to its computational characterization: relative to that lower-level interpretation, the classical "bits" are abstractions.  Treating the classical hardware level as a virtual machine recognizes this abstraction from the underlying quantum mechanical description of $\mathbf{D}$.  It also allows agnosticism about the measurement problem.  The details of the Hamiltonian $\mathcal{H}$ of any realistic macroscopic computer are in fact unknown, as are the details at the quantum level of the interactions between any macroscopic computer and its users or any other aspects of its environment.  The measurement interactions between user and device by which users identify and characterize the classical hardware level cannot, therefore, be specified in quantum-mechanical terms.  Treating the classical hardware level as a virtual machine, and hence treating the user-device interactions that specify it not as measurement but as interpretation, allows us to proceed with an analysis of user observations of and interactions with macroscopic states of $\mathbf{D}$ in the face of our nearly-complete ignorance of $\mathcal{H}$, and in particular in the face of our inability to strictly segregate and characterize those components of $\mathcal{H}$ that specifically enable the interpretation of the macroscopic behavior of $\mathbf{D}$ as the execution of some classical algorithm of interest.

\section{Data definition at the user-interface level}

Users use computers to run computations relevant to real or possible states of the world.  The fundamental semantic constraint on any computational device $\mathbf{D}$ is, therefore, that an interpretation $I$ exists such that the following diagram commutes:

\setlength{\unitlength}{1.0cm}
\begin{picture}(15,8)
\put(7.3,6){\vector(1,0){1.5}}
\put(7.3,3.2){\vector(1,0){1.5}}
\put(6.5,4){\vector(0,1){1}}
\put(9.6,4){\vector(0,1){1}}
\put(6.3,6){$S|_{t}$}
\put(9.3,6){$S|_{t + \Delta t}$}
\put(6.1,3){$D^{UI}_{i}$}
\put(9.3,3){$D^{UI}_{f}$}
\put(6,4.3){$I$}
\put(9.8,4.3){$I$}
\put(7.4,6.2){\textit{Process}}
\put(7.8,2.6){$A$}
\put(7.2,1){\textit{Diagram 1}}
\end{picture}

where $S|_{t}$ and $S|_{t + \Delta t}$ indicate states of some external system at some times $t$ and $t + \Delta t$, \textit{Process} is a dynamic or symbolic process acting on that system, $D^{UI}_{i}$ and $D^{UI}_{f}$ indicate initial and final contents of a data structure at the UI level, and $A$ is an algorithm that produces the final data from the initial data.  Commutivity of Diagram 1 for a large set of initial and final data values justifies treating $\mathbf{D}$ as a process virtual machine for $A$; in cases in which \textit{Process} is symbolic, it may be possible to prove that Diagram 1 commutes and hence prove that $A$ is correct.

The extension of this fundamental semantic constraint to lower-level virtual machines is well-understood \cite{tan76, smith05}; it requires an emulation mapping $E$ from the lower level virtual machine, and eventually from the classical hardware level $D^{H}$, such that the following diagram commutes:

\setlength{\unitlength}{1.0cm}
\begin{picture}(15,11)
\put(7.3,9){\vector(1,0){1.5}}
\put(7.3,6.2){\vector(1,0){1.5}}
\put(6.5,7){\vector(0,1){1}}
\put(9.6,7){\vector(0,1){1}}
\put(6.3,9){$S|_{t}$}
\put(9.3,9){$S|_{t + \Delta t}$}
\put(6.1,6){$D^{UI}_{i}$}
\put(9.3,6){$D^{UI}_{f}$}
\put(6,7.3){$I$}
\put(9.8,7.3){$I$}
\put(7.4,9.2){\textit{Process}}
\put(7.8,5.6){$A$}
\put(7.3,3.2){\vector(1,0){1.5}}
\put(6.5,4){\vector(0,1){1}}
\put(9.6,4){\vector(0,1){1}}
\put(6,4.3){$E$}
\put(9.8,4.3){$E$}
\put(6.1,3){$D^{H}_{i}$}
\put(9.3,3){$D^{H}_{f}$}
\put(7.5,2.5){\textit{Trace}}
\put(7.2,1){\textit{Diagram 2}}
\end{picture}

where \textit{Trace} is the classical hardware level execution trace of $A$.  The composition $I \circ E$ defines a semantic interpretation of the behavior of $\mathbf{D}$ at the classical hardware level $D^{H}$.  This interpretation assigns an external-world referent to every distinct virtual state traversed by \textit{Trace}, including in particular $D^{H}_{i}$ and $D^{H}_{f}$.

It is important to emphasize that \textit{Trace} is an interpretation of the behavior of $\mathbf{D}$, not a dynamical theory.  Some components of $\mathbf{D}$ that are essential to its function, including those involved in power distribution and heat dissipation, are systematically ignored in \textit{Trace}, which is defined at the level of description of a partial functional specification for $\mathbf{D}$'s hardware. The observations of $\mathbf{D}$'s behavior that verify compliance with \textit{Trace} are not observations of individual quantum states, but of uncharacterized ensembles of quantum states that are indistinguishable and hence equivalent at the spatial and temporal resolution of the observations.  It is a benefit of viewing the classical hardware level as a virtual machine emulated by $\mathbf{D}$ that the non-dynamic character of \textit{Trace} is made clear.  \textit{Trace} looks like and is often thought of as a bottom-up description of \textit{how} $\mathbf{D}$ implements $A$, but in fact \textit{Trace} is a top-down specification of a virtual machine that is driven by interpretability requirements set at the UI level.

\section{Data definition for quantum registers}

Quantum computers are distinguished by their use of controlled manipulations of entangled states of physical devices implementing qubits to execute computational processes on information encoded in the global state of a qubit array implementing a quantum register $\mathbf{R}$ \cite{galindo01}.  Such quantum-scale manipulations, including the manipulations involved in preparing the initial entangled state of $\mathbf{R}$, clearly cannot commute with observations of the states of the individual qubits.  Without the in-principle ability to observe individual qubit states at any stage of the computation, is not possible to define \textit{Trace} in the sense required by Diagram 2 at the scale of quantum encoding.  Hence quantum-scale encodings cannot inherit their external-world referents from an emulation mapping, even one defined, as in Diagram 2 and typical practice, only at the initial and final steps of the computation.  The external-world referents of quantum-scale encodings must, therefore, be stipulated.  As only the final states of $\mathbf{R}$ are directly observable, the fundamental stipulation is of a mapping $M_{f}$ from any observable final state of $\mathbf{R}$ to some classical data item at the UI level, i.e.

\begin{equation}
M_{f}: \mathbf{O}|R_{\mathit{j}}> \longmapsto \mathit{D^{UI}_{j}}
\end{equation}

where the observable $\mathbf{O}$ is chosen as part of the definition of the quantum algorithm $\mathbf{Q}$, and the virtual state ${D^{UI}_{j}}$ is the UI level representation of the intended external referent of the global state $|\mathbf{R}_{\mathit{j}}>$.  This stipulated final-state interpretation mapping induces a data definition mapping $M_{i}$ from $\mathbf{R}$ to $D^{UI}$.  A classical representation of this data definition must be accessible from the UI level to allow users both to interpret the qubit states in terms of their external referents once they are measured, and to evoke the quantum computation from the appropriate steps in a classical program, i.e. to pass data to the quantum computation through an API.  The commutativity requirement relating the stipulation of external referents of a quantum register $\mathbf{R}$ by the data definition mapping $M_{i}$, the post-computation measurement of the final states of $\mathbf{R}$ by the measurement operator $\mathbf{O}$, and the post-measurement interpretation $M_{f}$ is shown in Diagram 3.  Replacing the bottom square of Diagram 2 with Diagram 3 fully describes the UI level semantics of $\mathbf{Q}$.

\setlength{\unitlength}{1.0cm}
\begin{picture}(15,8)
\put(7.3,6){\vector(1,0){1.5}}
\put(7.3,3.1){\vector(1,0){1.5}}
\put(6.5,4){\vector(0,1){1}}
\put(9.6,4){\vector(0,1){1}}
\put(6.3,6){$D^{UI}_{i}$}
\put(9.3,6){$D^{UI}_{f}$}
\put(6.1,3){$|\mathbf{R}_{\mathit{i}}>$}
\put(9.2,3){$|\mathbf{R}_{\mathit{f}}>$}
\put(5.7,4.3){$M_{i}$}
\put(9.8,4.3){$\mathbf{O} \circ \mathit{M_{f}}$}
\put(7.8,6.2){$A$}
\put(7.8,2.6){$\mathbf{Q}$}
\put(7.2,1){\textit{Diagram 3}}
\end{picture}

Three cases exhaust the possibilities for stipulating the external referents of a quantum register by a data-definition mapping $M_{i}$: 1) the states of the qubits in the computational basis $\{|0>, |1>\}$ can directly encode the external referents by a well-defined encoding algorithm; 2) external referents can be arbitrarily assigned to the states of the qubits in the computational basis by an arbitrary explicit mapping; 3) external referents can be arbitrarily assigned to the global states in the computational basis of the qubit register $\mathbf{R}$ by an arbitrary explicit mapping.  In the first two cases, the external referents of the global states of $\mathbf{R}$ are inherited via a classical algorithm from the external referents assigned to the individual qubit states; in the third case no such inheritance is required.  In each of these cases, the measurement operator $\mathbf{O}$ employed at the end of the computation, or in measurement-based computations at the end of each qubit's role in the computation \cite{briegel09}, returns real values of squared qubit amplitudes.  The post-measurement interpretation mapping $M_{f}$ relates these squared amplitudes in the computational basis to classical user-interpretable data values $D^{UI}_{f}$.  An $M_{f}$ that maps the squared qubit amplitudes to real numbers acts as an analog interpretation of the computation; an $M_{f}$ that thresholds the amplitudes to zero or one and maps the resulting binary vector to a discrete data item acts as a digital interpretation of the computation.  The classical-side computational costs associated with the mappings $M_{i}$ and $M_{f}$ in each of these cases will be considered, with examples, below.

\subsection{Case 1: Direct encoding by qubit basis states}

A direct encoding of the external referents of a computation by the basis states of the individual qubits requires an isomorphism from the external domain to the ordered array of binary basis states.  The external referents being assigned must, therefore, be isomorphic to an ordered array of binary numbers, i.e. they must be isomorphic to a binary representation of the integers.  Hence Case 1 can be regarded, without loss of generality, as the case in which the external referents of the computation are the integers.

Shor's factoring algorithm \cite{shor94, shor97} is the canonical exemplar of this case.  The $n$ qubits composing $\mathbf{R}$ directly encode the integers from $0$ to $2^{n} - 1$; no explicit assignment of values to the qubits needs to be made.  The Shor algorithm Hamiltonian $\mathbf{Q}_{\mathit{Shor}}$ ideally amplifies the amplitudes of the qubits encoding the most probable factors to unity while supressing all others to zero; hence the final-state interpretation $M_{f}$ for Shor's algorithm is digital.  The addresses of the final qubit states corresponding to most probable factors are mapped by $M_{f}$ to the integers that they represent simply by interpreting the qubits as binary representations, using the natural ordering of the qubits inherited from $\mathbf{R}$.  Hence the UI needs only represent that the encoding is binary and that the qubits are to be read in the order in which they appear in $\mathbf{R}$ to inform the user of the decoding.    

All quantum algorithms in Case 1 can be interpreted as acting on the integers, so both initial data definition and final post-observation interpretation for all such algorithms can be handled in the same way as for Shor's algorithm.  Measurements reporting continuous values are not relevant for such algorithms.  The computational cost of data definition in Case 1 is, therefore, negligible: it consists solely of informing the user of the binary coding.  With appropriate UI level conventions, this cost can be reduced to one bit.

\subsection{Case 2: Arbitrary assignment to qubit basis states}

User interface level data items $\{x_{i}\}$ corresponding to arbitrary external referents can be assigned to the $n$ individual qubits plus the $NULL$ state $|0,0, ... 0>$ of $\mathbf{R}$ provided the UI level representation $x$ of any given external referent within the domain of interest can be generated by a classical algorithm $A$ as $x = A(x_{i})$.  This condition is satisfied in general by problem domains in which the external referents of interest are values in the range of some mathematical function $f(x_{i})$; it is also satisfied by axiomatic inferential systems.

The canonical exemplar of Case 2 is quantum simulation \cite{lloyd96, buluta09, brown10}.  The $n$ basis vectors $\{|x_{i}>\}$ of a quantum system of interest $S$ are assigned arbitrarily to the $n$ individual qubits of $\mathbf{R}$; the zero state vector of $S$ is assigned to the $NULL$ state of $\mathbf{R}$.  A quantum algorithm $\mathbf{Q}_{\mathit{S}}$ is then chosen so that Diagram 4 commutes, where $\mathcal{H}_{\mathit{S}}$ is the Hamiltonian of $S$ and $A_{S}$ is a classical algorithmic model of $S$, e.g. a Schr\"odinger equation.

\setlength{\unitlength}{1.0cm}
\begin{picture}(15,11)
\put(7.3,9){\vector(1,0){1.5}}
\put(7.3,6.2){\vector(1,0){1.5}}
\put(6.5,7){\vector(0,1){1}}
\put(9.6,7){\vector(0,1){1}}
\put(6.3,9){$S|_{t_{i}}$}
\put(9.3,9){$S|_{t_{f}}$}
\put(6.1,6){$D^{UI}_{i}$}
\put(9.3,6){$D^{UI}_{f}$}
\put(6,7.3){$I$}
\put(9.8,7.3){$I$}
\put(7.7,9.2){$\mathcal{H}_{\mathit{S}}$}
\put(7.7,5.6){$A_{S}$}
\put(7.3,3.1){\vector(1,0){1.5}}
\put(6.5,4){\vector(0,1){1}}
\put(9.6,4){\vector(0,1){1}}
\put(5.8,4.3){$M_{i}$}
\put(9.8,4.3){$\mathbf{O_{\mathit{S}}} \circ \mathit{M_{f}}$}
\put(5.9,3){$|\mathbf{R}_{\mathit{i}}>$}
\put(9.1,3){$|\mathbf{R}_{\mathit{f}}>$}
\put(7.7,2.5){$\mathbf{Q}_{\mathit{S}}$}
\put(7.2,1){\textit{Diagram 4}}
\end{picture}

The analog final-state interpretation $M_{f}$ maps the measured squared amplitude of the $i^{th}$ qubit to the squared coefficient of the basis vector $|x_{i}>$ to which the $i^{th}$ qubit is mapped by $M_{i}$.  Because the assignment of qubits to basis vectors is arbitrary, $M_{f}$ must effectively perform a look-up operation on the association table defined by $M_{i}$; there are $n$ items in this table, so the cost of this final mapping step is $O(n)$.

The simulation of any linear system can clearly be treated as a quantum simulation, and hence incurs a data-definition cost that is $O(n)$.  Nonlinear and inferential systems cannot be so treated; for such systems the final-state qubit amplitudes have no direct interpretation as linear coefficients of the generators $x_{i}$.  In this case, $M_{f}$ is a nonlinear digitizing function of the qubit amplitudes that maps an arbitrary global state $|\mathbf{R}_{\mathit{f}}>$ in the computational basis to a discrete referent $A(x_{i})$, i.e.

\begin{equation}
M_{f}: |\mathbf{R}_{\mathit{f}}> = \mathit{\sum \lambda_{j} |j> \longmapsto A(x_{i})}.
\end{equation}

Reversing the initial assignment $M_{i}$ of external referents to qubits, Eq. 4 is just:

\begin{equation}
A: x = \sum \lambda_{j} |x_{j}> \longmapsto A(x).
\end{equation}

Hence in this case, the final-state interpretation map $M_{f}$ is merely a notational variant of the classical algorithm $A$.  For nonlinear and inferential systems with Case 2 data definition mappings, therefore, final state interpretation has the same computational cost as the classical algorithmic model $A$, so quantum computation is pointless.

\subsection{Case 3: Arbitrary assignment to global states of $\mathbf{R}$}

The most general option for data definition is to arbitrarily assign external referents to the $N$ global states of $\mathbf{R}$ in the computational basis.  Quantum computation is then initiated in a fully-entangled state in which every qubit is prepared with the same amplitude.

The canonical exemplar of this case is the Grover algorithm \cite{grover96, grover97} for searching an unordered database.  In the reverse telephone-directory search application envisioned by Grover \cite{grover96, grover97}, quantum search of an unordered database of $N$ telephone numbers is employed to rapidly (in $O(\sqrt{N})$ time) retrieve a key into a classical database containing subscriber names and addresses, thus avoiding an $O(N)$ search of the classical database.  Each telephone number is arbitrarily assigned to a global state of $\mathbf{R}$, which is prepared in an initial global state with equal amplitudes $(\frac{1}{\sqrt{N}}, \frac{1}{\sqrt{N}}, ..., \frac{1}{\sqrt{N}})$.  The Grover algorithm is implemented by an operator $\mathbf{Q}_{\mathit{Grover}}$ that selectively amplifies the global state meeting a given search criterion; at the end of the search this amplified global state is interpreted by $M_{f}$ as a telephone number.  Because the telephone numbers have been assigned to global states of $\mathbf{R}$ arbitrarily, final-state interpetation requires a look-up operation on the initial assignment table, i.e. a classical search of a classical data structure such as:

 \begin{center}
\setlength{\unitlength}{10mm}
\begin{picture}(6,7)
  \put(0,5){\line(1,0){6}}
  \put(0,4){\line(1,0){6}}
  \put(0,3){\line(1,0){6}}
  \put(0,2){\line(1,0){6}}
  \put(.1,5.3){$|1>$}
  \put(1.2,5.3){$415-389-1133$}
  \put(4.6,5.3){$uid~1$}
  \put(.1,4.3){$|2>$}
  \put(1.2,4.3){$415-486-6129$}
  \put(4.6,4.3){$uid~2$}
  \put(.1,3.3){$|3>$}
  \put(1.2,3.3){$415-492-0206$}
  \put(4.6,3.3){$uid~3$}
  \put(.3,2.3){$...$}
  \put(2.5,2.3){$...$}
  \put(4.9,2.3){$...$}
  \put(.1,1.3){$|N>$}
  \put(1.2,1.3){$415-472-6775$}
  \put(4.6,1.3){$uid~N$}
\thicklines
  \put(0,1){\line(0,1){5}}
  \put(1,1){\line(0,1){5}}
  \put(4.5,1){\line(0,1){5}}
  \put(6,1){\line(0,1){5}}
  \put(0,6){\line(1,0){6}}
  \put(0,1){\line(1,0){6}}
\end{picture}
\end{center}
\begin{quote}
Figure 1:  Classical data structure relating global qubit register states $|1>$ through $|N>$ of a device executing the Grover algorithm (column 1) to telephone numbers (column 2) and their associated unique identifiers in a classical database (column 3).
\end{quote}
 
where classical unique identifiers are included as classical database keys.  This classical search requires $O(N)$ steps; indeed it is the very search that the Grover algorithm is intended to replace.

Nothing in the above analysis depends on the details of $\mathbf{Q}_{\mathit{Grover}}$; it applies to any quantum algorithm with a Case 3 data definition and a digitizing measurement of the final state.  Analog measurements, however, have no application in Case 3.  Because the external referents are assigned by $M_{i}$ to global states in the computational basis, mixed states in this basis would correspond to mixtures of the classical referents, which have no classical meaning.  No quantum algorithm with a Case 3 data definition can, therefore, offer meaningful speed-up compared to classical computation; the cost of final-state interpretation is always $O(N)$.

\section{Data definition for quantum instructions}

While ``quantum data, classical control" may be considered from a first-principles viewpoint one of many approaches to quantum computation \cite{selinger04}, any practical quantum computer will require a layer of control specified at the classical UI level.  Users at minimum require the ability to select which quantum algorithm to execute (or to call via an API from within a classical program) and to specify the values of any parameters that control the quantum algorithm's operation.  The Grover algorithm again provides an illustrative example.  Grover search amplifies a global state $|i>$ of the quantum register $\mathbf{R}$ that is ``marked" by uniquely satisfying a search condition $C(|k>)$ such that $C(|i>) = 1$ and $C(|j>) = 0$ for all $j \neq i$ \cite{grover96, grover97}.  The condition $C(|k>)$ is specified by the user; in reverse telephone directory search, it corresponds to the telephone number to search for.  The value of the index $i$ corresponding to a given telephone number must be looked up in a classical data structure such as the table shown in Fig. 1; this classical search requires $O(N)$ steps, as is clear from physical implementations of the Grover algorithm (e.g. the $n = 4$ ion-trap implementation of Brickman et al. \cite{brickman05}, where $i$ is specified by the values of the UI level parameters $\alpha$ and $\beta$).  Hence specifying the parameters for the Grover algorithm requires solving the problem the Grover algorithm is designed to solve.  This is in stark contrast to the situation with Shor's algorithm, in which the only parameter that must be passed at the UI or API level is the number to be factored.

In general, the cost of specifying a quantum algorithm to execute and its run-time parameters may be characterized in the Quantum Turing Machine (QTM) model \cite{deutsch85}.  A QTM employs two quantum registers: the $n$-qubit data register $\mathbf{R}$ as described above, and an $m$-qubit program register $\mathbf{P}$ that specifies the quantum algorithm $\mathbf{Q}$ to be applied to the data together with the values of all relevant run-time parameters.  The specification of any non-trivial quantum algorithm and its parameter values for a specific run is arbitrary with respect to the physical addresses of the qubits composing $\mathbf{P}$; the values of the $M = 2^{m}$ global states of $\mathbf{P}$ must, therefore, be explicitly stipulated at the UI (equivalently, the API) level of a QTM.  Hence $M$ classical steps are required to specify a quantum algorithm and its run-time parameters to a QTM.  The equivalent specification to a practical quantum computer that incorporates an API allowing a fixed set of quantum algorithms $\{\mathbf{Q}_{\mathit{i}}\}$ to be called by name will be $M^{\prime}_{j} + 1$, where $M^{\prime}_{j} < M$ is the number of independent run-time parameters of $\mathbf{Q}_{\mathit{j}}$.

The considerations of Sect. 4 limit practical questions concerning the maximum value of $M^{\prime}$ to the cases of quantum algorithms defined over the integers and quantum simulators.  In the former case, $M^{\prime}$ counts the number of integer inputs to the computation; this number can be assumed to be small for ``interesting" computations.  
In the case of quantum simulators, the $M^{\prime}$ run-time parameters correspond to user-specifiable boundary conditions on the simulation.  In principle, these could be arbitrarily complex.  A quantum simulation will offer meaningful speed-up over classical computation if $M^{\prime}$ is $O(n)$ or less; quantum simulation is pointless if $M^{\prime}$ is $O(N)$.  The boundary conditions of the simulation cannot, therefore, depend arbitrarily on the global state of $\mathbf{R}$, they must be functions of the $n$ basis vectors only.  To be implemented by unitary operators acting on the qubits composing $|\mathbf{R}>$, they must be linear functions of these basis vectors.

\section{Conclusions}

The above analysis limits the practical utility of quantum algorithms to two cases: algorithms defined over the integers, and algorithms that simulate linear systems with linear boundary conditions.  The cost of data definition in the former case can be reduced to 1 bit plus the cost of specifying inputs; in the latter case it is $O(n)$.  In all other cases, the computational cost of data definition for $\mathbf{R}$ plus the cost of specifying run-time parameters equals or exceeds the computational cost of a classical algorithm for the task under consideration.  

The costs of data definition for $\mathbf{R}$ derived here follow solely from the lack of commutativity between the local Hamiltonian $\mathbf{Q}$ and the final-state observable $\mathbf{O}$ that forces the stipulative semantics of Diagram 3.  Without emulation mappings based on well-defined execution traces, stipulative semantics require final-state data interpretation.  The costs of this post-measurement interpretation step have not been adequately considered in previous analyses of quantum algorithms.  The present results show that they are significant.

These considerations suggest that quantum computation may be of limited practical utility not because it is hard to implement, but because it is hard to interface to users with classical problem definitions and a requirement for classical data representations at the UI level.  User requirements for classical data impose a classical algorithmic overhead on quantum computing systems.  This overhead is small only for problems with natural mappings to the computational basis of a quantum computer.

\section*{Statement regarding conflict of interest}

The author states that he has no financial or other conflicting interests in any results reported in this paper.


\begin{thebibliography}{99}

\bibitem{tan76} Tanenbaum, A. S. 1976 \textit{Structured Computer Organization.}  Upper Saddle River, NJ: Prentice Hall.

\bibitem{smith05} Smith, J. E., Nair, R. 2005 \textit{Virtual Machines: Versatile Platforms for Systems and Processes.}  San Francisco: Morgan Kauffman.

\bibitem{aaronson05} Aaronson, S. 2005 NP-complete problems and physical reality.  \textit{ACM Sigact News} 36, 30-52.  arXiv quant-ph/0502.072v2.

\bibitem{galindo01} Galindo, A. \& Martin-Delgado, M. A.  Information and Computation: Classical and quantum aspects.  \textit{Rev. Mod. Phys.} 74, 347-423 (2002).

\bibitem{briegel09} Briegel, H. J., Browne, D. E., D\"ur, W., Raussendorf, R. \& van den Nest, M.  Measurement-based quantum computation.  \textit{Nat. Phys.} \textbf{5}, 19-26 (2009).

\bibitem{shor94} Shor, P. W.  Algorithms for quantum computation: Discrete logarithms and factoring.  \textit{Proc. 35th Ann. Symp. Found. Comp. Sci.} IEEE Computer Society, 124-134 (1994).

\bibitem{shor97} Shor, P. W.  Polynomial-time algorithms for prime factorization and discrete logarithms on a quantum computer.  \textit{SIAM J. Comput.} \textbf{26}, 1484-1509 (1997).

\bibitem{lloyd96}  Lloyd, S.  Universal quantum simulators.  \textit{Science} \textbf{273}, 1073-1078 (1996).

\bibitem{buluta09}  Buluta, I. \& Nori, F.  Quantum simulators.  \textit{Science} \textbf{326}, 108-111 (2009).

\bibitem{brown10}  Brown, K. L., Munro, W. J. \& Kendon, V. M.  Using quantum computers for quantum simulation.  arXiv:quant-ph/1004.5528v1 (2010).

\bibitem{grover96} Grover, L. K.  A fast quantum mechanical algorithm for database search.  \textit{Proceedings of the ACM Symposium on Theory of Computing} (pp. 212-219) Philadelphia, PA (1996).

\bibitem{grover97} Grover, L. K. Quantum mechanics helps in searching for a needle in a haystack.  \textit{Phys. Rev. Lett.} \textbf{79}, 325-328 (1997).

\bibitem{selinger04} Selinger, P.  Toward a quantum programming language.  \textit{Math. Struct. Comp. Sci.} \textbf{14}, 527-586 (2004).

\bibitem{brickman05} Brickman, K.-A., Haljan, P. C., Lee, P. J., Acton, M., Deslauriers, L. \& Monroe, C. (2005) Implementation of Grover's quantum search algorithm in a scalable system.  \textit{Physical Review A} \textbf{72}, 050306(R).

\bibitem{deutsch85} Deutsch, D.  Quantum theory, the Chruch-Turing principle and the universal quantum computer.  \textit{Proc. Royal Soc. London A} \textbf{400}, 97-117 (1985).


\end{thebibliography}
\end{document}